\documentclass[12pt]{iopart}

\usepackage{epsfig}
\usepackage{amsfonts, amssymb}
\usepackage[usenames]{color}
\usepackage{multirow}
\usepackage[applemac]{inputenc}
\usepackage{subfigure}

\newcommand\eq[1]{(\ref{#1})}

\begin{document}
\setlength{\unitlength}{1mm}

\title[]{Using CMB data to constrain non-isotropic Planck-scale modifications to Electrodynamics 
}

\author{Giulia Gubitosi}
\address{ Berkeley Lab \& University of California, Berkeley, 
CA 94720, USA}
\author{Marina Migliaccio}
\address{Universit\`a di Roma Tor Vergata,Via della Ricerca Scientifica,1, Roma, Italy}
\author{Luca Pagano}
\address{Jet Propulsion Laboratory, California Institute of Technology, 4800 Oak Grove Drive, Pasadena, California, U.S.A.}
\author{Giovanni Amelino-Camelia, Alessandro Melchiorri}
\address{Dipartimento di Fisica, Universit\`a La Sapienza, P. le A. Moro 2, Roma, Italy  \\ and Sezione Roma1 INFN, P.le Aldo Moro 2, 00185 Rome, Italy}
\author{Paolo Natoli}
\address{Dipartimento di Fisica, Università di Ferrara, Via G. Saragat 1, Ferrara\\
Agenzia Spaziale Italiana Science Data Center, c/o ESRIN, via Galileo Galilei, Frascati, Italy \\
INAF/IASF Bologna, Via Gobetti 101, Bologna, Italy}
\author{Gianluca Polenta}
\address{Agenzia Spaziale Italiana Science Data Center, c/o ESRIN, via Galileo Galilei, Frascati, Italy\\
INAF Osservatorio Astronomico di Roma, via di Frascati 33, 00040 Monte Porzio Catone, Italy}

\begin{abstract}
We develop a method to constrain non-isotropic features of Cosmic Microwave Background (CMB) polarization, of a type expected to arise in some models describing quantum gravity effects on light propagation.
We describe the expected signatures of this kind of anomalous light propagation on CMB photons, showing that it will produce a non-isotropic birefringence effect, i.e. a rotation of the CMB polarization direction whose observed amount  depends in a peculiar way on the observation direction.
We also show that the sensitivity levels expected for CMB polarization studies by the \emph{Planck} satellite are sufficient for testing these effects if, as assumed
 in the quantum-gravity literature, their magnitude is set by the minute Planck length.
\end{abstract}

\maketitle

\section{Introduction}

One of the main open problems in modern physics is finding
a description for phenomena taking place in the so-called ''quantum-gravity realm'',
at  scales where both gravitational and quantum effects are non-negligible \cite{Stachel:1999we,Carlip:2001wq}.
Several alternative candidate theories are being developed but discriminating among them
experimentally is extremely difficult as a result of the smallness of
the characteristic quantum-gravity length scale,
the Planck length $L_{p}\sim 10^{-35}m$
(which of course corresponds to an energy scale,
the Planck scale $E_{p}\sim 10^{28}eV$, which is huge by particle-physics standards).
Nonetheless it is now well established that astrophysical observations  can provide significant information about physics at the Planck scale \cite{AmelinoCamelia:2008qg}, and recently it has been pointed out that also  cosmological observations can be helpful. The opportunities on the cosmology side are becoming more mature as we start to have high precision data, and in particular it was recently shown that current data on Cosmic Microwave Background  (CMB) radiation  have already enough sensitivity to be used to establish
meaningful constraints on
the effects predicted by some Planck-scale
theories \cite{Gubitosi:2009eu, systematics, birefringenceproceeding}.


This is possible because, even if CMB radiation is characterized by quite low energies, it has been propagating for very long times, so that it is potentially  subject to a large accumulation of new physics effects affecting photons propagation.
In \cite{Gubitosi:2009eu} some of us were able to show that measurable changes can be induced on CMB polarization power spectra due to anomalous light propagation, of the kind predicted by some descriptions of modified
electrodynamics \cite{Myers:2003fd,Gleiser:2001rm}
which are much in use in the literature as effective description of the implications
for electrodynamics of the Planck-scale (possibly quantum) structure of
spacetime. These anomalous properties of light lead to a birefringent behavior of CMB radiation, producing a rotation of its polarization direction, which is detectable through an analysis of the cross-correlation power spectra
\cite{Lepora:1998ix,Lue:1998mq,Feng:2006dp,Xia:2009ah,Kahniashvili:2008va}.

%
%
%
In \cite{Gubitosi:2009eu} anomalous light propagation was encoded in the
 deformed-electrodynamics
model introduced by Myers and Pospelov \cite{Myers:2003fd},
a dimension-five effective field theory coupling the electromagnetic field with an external fixed four-vector, which explicitly breaks Lorentz symmetries.
This model was also tested  in several astrophysical contexts, leading to very stringent constraints \cite{Galaverni:2007tq, Maccione:2008iw, Gleiser:2001rm, Jacobson:2002ye, AmelinoCamelia:2002dx}.
In \cite{anysotropy} it was observed that these bounds exploit
significantly the spatial isotropy regained by the {\it ad hoc}
choice of having a purely timelike symmetry-violating four-vector, and actually this choice is only available for a
restricted class of frames of reference,
since the four-vector  will of course still acquire a spatial component
in other boosted frames. This assumption can be limiting in two ways.
Constraints on one single component of a four vector derived in different reference frames (like the rest frame of two different astrophysical sources)  cannot be compared without any information on the other components of the four-vector. Moreover in \cite{anysotropy} it was shown that the limits placed assuming the four-vector to be purely timelike do not give reliable information on the most general case in which the four-vector has all the components different from zero, so that also space isotropy is violated.

The success of the attempts to constrain isotropic anomalous light propagation
through CMB observations provide motivations
to  investigate more widely how Planck-scale modifications of electrodynamics
can affect propagation of light in a way that would be detectable through
cosmological observations.

In this paper we describe the expected behavior of CMB photons in presence of the non-isotropic  Lorentz-symmetry violations described by the most general version of the model proposed in Ref.~\cite{Myers:2003fd}.
We  show that  anomalies  can affect CMB radiation propagation, in a way that can be described as non-isotropic birefringence, i.e. a rotation of the polarization direction  whose amount is characterized by a peculiar dependence on the observation direction.

The most relevant characteristic of the model described in this paper is that anomalous light behavior depends on its propagation direction with respect to a preferred direction codified within
the symmetry-breaking vector $n_{\alpha}$. To make more explicit the difference with  the previously studied case \cite{Myers:2003fd, Gubitosi:2009eu} in which space isotropy is preserved ($n_{\alpha}=(n_{0},0,0,0)$), we will specialize at the end of section 2 to the complementary case in which the symmetry breaking vector has only the space components different from zero.

We  also  show that in this case the peculiar space-dependence of the birefringence effect  is such that the rotation can not be detected exploiting the standard tool of correlation power spectra, which instead was shown to be valuable for constraining the isotropic rotation effect studied in \cite{Gubitosi:2009eu}.

So we need to develop a specific method to detect such anisotropic effects, and this is described in section 3.
In section 4 we forecast on the sensitivity level
that will be reachable with the \emph{Planck} satellite observations, showing that it will allow to constrain the model even beyond the Planck scale level.

\section{CMB photons propagation in presence of non-isotropic Lorentz symmetry violations}\label{sec:birefringence}

In  \cite{anysotropy} it is described the  behavior of photons in presence of non-isotropic violations of Lorentz invariance,  resulting from the generalization of the Myers-Pospelov model, with nonzero spatial components of the four-vector.
The Lagrangian density of the model is:
\begin{equation}
\mathcal L_{QG}=-\frac{1}{4}F_{\mu\nu}F^{\mu\nu}
 +\frac{1}{2E_P} n^\alpha F_{\alpha\delta}n^\sigma
  \partial_\sigma(n_\beta\varepsilon^{\beta\delta\gamma\lambda}F_{\gamma\lambda})
  ~, \label{eq:lagrangianJOC}
\label{eq:lagrangianMP}
\end{equation}
where $n_{\alpha}$ is the symmetry-breaking four-vector and the coupling constant between the electromagnetic field and the vector is given by the inverse of the Planck energy, so setting the appearance of new physics at that scale.

The relevant equations for the dispersion relation and the eigenstates of the field propagation resulting from the above Lagrangian are \cite{anysotropy} :
\begin{equation}
\omega_{\pm} \simeq |\vec p|\pm \frac{1 }{E_{p}} |\vec p|^2 \left( n_0
 + \frac{{\vec n}\cdot {\vec p}}{|\vec p|}\right)^3 \label{eq:disprel}
\end{equation}

\begin{equation}
\vec{\mathcal E}_\pm =
 \left(
\begin{array}{c}
 \mp\frac{2}{E_{p}}\frac{|\vec n|(\vec p\cdot \vec n+|\vec p|n_0)^2 |\vec p-(\vec p\cdot\hat n)\hat n|}{\sqrt{2}|\vec p|^2}\\
\pm \frac{i }{\sqrt{2}}+\frac{i }{\sqrt{2}E_p}\frac{|n|^2 |\vec p-(\vec p\cdot\hat n)\hat n|^2 (\vec p\cdot \vec n+|\vec p|n_0)}{|\vec p|^2}\\
\frac{1}{\sqrt{2}}\mp\frac{1}{\sqrt{2}E_p}\frac{|\vec n|^2 |\vec p-(\vec p\cdot\hat n)\hat n|^2 (\vec p\cdot \vec n+|\vec p|n_0)}{|\vec p^2|}
\end{array}
\right)\label{eq:eigenstates}
\end{equation}
where the field eigenstates are written in the basis
\begin{equation}\label{eq:baseBIS}
 \left\{\frac{\vec p}{|\vec p|},\frac{ \hat n\times \vec p}{\sqrt{p^2
 -(\hat n\cdot \vec p)^2}}, \frac{-\vec p (\vec p\cdot \hat n)
+\hat n |\vec p|^2}{|\vec p| \sqrt{p^2-(\hat n\cdot \vec p)^2}}\right\}~,
\end{equation}
such that the first component of the field is the longitudinal one and the other two components describe the field in the plane transverse to the propagation direction.
\footnote{We use the notation of Jones three-dimensional
vectors \cite{anysotropy}:  the polarization state of the field $$\vec E(x,t)=Re\left[\left(E_x \hat x+E_y \hat y
+E_z\hat z  \right)e^{i(\vec k\cdot x-\omega t)}\right]$$

with $E_x,E_y,E_z$ complex numbers is represented as

 \begin{equation}
  \vec{\mathcal E}=\frac{1}{\sqrt{\mathcal E_x{}^2+\mathcal E_y{}^2+\mathcal E_z{}^2}}\left(\begin{array}{c}
 \mathcal {E}_x \\
 \mathcal {E}_y \\
 \mathcal {E}_z
\end{array}\right).
\end{equation}}

It is then clear that the model leads to the emergence of several anomalous effects, whose intensity is always depending on the relative orientation between the symmetry-breaking four vector $n_{\alpha}$ and the field wave-vector. Eq.\eq{eq:eigenstates} shows that the field eigenstates are elliptical, and moreover are not transverse, having a Planck-scale-suppressed longitudinal component.  Eq.\eq{eq:disprel} describes a modification of the dispersion relation, of opposite sign for the orthogonal polarization states of the field $\vec{\mathcal E}_\pm $ (this is a birefringent behavior), and modulated by the angle between the spatial part of $n_{\alpha}$ and the field propagation direction $\vec p$.


All these anomalous properties are of course strongly suppressed by the Planck scale,
but in the following we will show that  some of them can be indeed amplified by very long propagation times, compensating for the Planck-scale suppression, so that they can produce observably large effects.



%
\subsection{Effective behavior of CMB radiation}
CMB photons have  a partial linear polarization due to their Thomson scattering with electrons at the last scattering surface in presence of a quadrupolar anisotropy. After this last scattering they propagate (almost) freely toward us.

To study photon propagation described by Eqs. \eq{eq:disprel} and \eq{eq:eigenstates} we would like to write a plane wave with the known polarization properties of CMB  as a linear combination of the eigenstates  $\vec{ \mathcal E}_{\pm}$ of Eq. \eq{eq:eigenstates}. But a  transverse linearly polarized field cannot be written as a linear combination of these two eigenstates, meaning that it does not belong to the solutions of the modified Maxwell equations following from the non-isotropic generalization of the  Myers-Pospelov model.
To overcome this difficulty, taking into account the fact that what we actually observe within the sensitivities presently available is that CMB is transverse and linearly polarized, we will assume that the original field is a solution of the modified Maxwell equations, which can be written as an expansion in powers of $\frac{1}{E_{p}}$, and at zeroth order can be described as a transverse linearly polarized state. So the field will be of the form

\begin{equation}
\label{eq:CMBfield}
\vec E=\vec E^{(0)}+\frac{1}{E_p}\vec E^{(1)}
\end{equation}
where $\vec E^{(0)}$ is transverse and linearly polarized and  $\vec E^{(1)}$ must be such that $\vec  E$ is a  solution of the modified Maxwell equations (to the first order in $\frac{1}{E_{p}}$), \emph{i.e.} the polarization state $\vec{\mathcal E}$ of $\vec E$ can be written as  a linear combination  of the eigenstates \eq{eq:eigenstates}:
\begin{equation}\label{eq:LinearCombination}
\vec{\mathcal E}\equiv \vec{\mathcal E}^{(0)}+\frac{1}{E_{p}}\vec{\mathcal E}^{(1)}=A \vec{ \mathcal E}_{+} +B\vec{ \mathcal  E}_{-}.
\end{equation}
 Of course  also the complex coefficients $A$ and $B$ can be expanded as a series in powers of $\frac{1}{E_p}$. To the first order: $A=A^{(0)}+\frac{1}{E_p}A^{(1)}, B=B^{(0)}+\frac{1}{E_p}B^{(1)}$.

 To enforce Eq. \eq{eq:LinearCombination} and constrain the form of $\vec{\mathcal E}^{(0)}$, $\vec{\mathcal E}^{(1)}$, $A$ and $B$, it is convenient to rewrite
 also the eigenstates in Eq. \eq{eq:eigenstates} pointing out the expansion in powers of $\frac{1}{E_{p}}$ (they were already written only up to the first order in $\frac{1}{E_{p}}$):
\begin{equation}\label{eq:eigenstatesSeries}
\vec{ \mathcal E_\pm}= \vec{\mathcal E_\pm^{(0)}}+\frac{\vec{\mathcal E_\pm^{(1)}}}{E_p} \equiv\left(
\begin{array}{c}
0  \\
 \pm \frac{i}{ \sqrt{2}}   \\
 \frac{1}{\sqrt{2}}
\end{array}
\right)+\frac{1}{E_p}\left(
\begin{array}{c}
\mp 2\frac{|\vec n|(\vec p\cdot \vec n+|\vec p|n_0)^2 |\vec p-(\vec p\cdot\hat n)\hat n|}{\sqrt{2}|\vec p|^2}  \\
 \frac{i }{\sqrt{2}}\frac{|n|^2 |\vec p-(\vec p\cdot\hat n)\hat n|^2 (\vec p\cdot \vec n+|\vec p|n_0)}{|\vec p|^2} \\
 \mp\frac{1 }{\sqrt{2}}\frac{|n|^2 |\vec p-(\vec p\cdot\hat n)\hat n|^2 (\vec p\cdot \vec n+|\vec p|n_0)}{|\vec p|^2}
\end{array}
\right).
\end{equation}
We set $t=0$ at the last scattering surface. At this time, the zeroth order of the CMB field in Eq. \eq{eq:CMBfield} is a linearly polarized transverse field:
$$\vec E^{(0)}(\vec r,t= 0)=\Re\{\left(\begin{array}{c}0\\a_1\\ a_2 \end{array}\right) e^{i \vec p\cdot \vec r}\}$$with $a_{1}$ and  $a_{2}$ real coefficients (we are still using the  basis \eq{eq:baseBIS}).
Asking   equation Eq. \eq{eq:LinearCombination} to be satisfied to the zeroth order in $\frac{1}{E_p}$, we find
\begin{equation}\label{eq:A0B0}
 \left\{\begin{array}{ccc}A^{(0)}&=&\frac{ a_2-i a_1}{\sqrt{2}}\\
B^{(0)}&=&\frac{a_2+i a_1}{\sqrt{2}}
        \end{array}\right.
\end{equation}
which is indeed the relation between the components of a linearly polarized field written in a linear basis and the components of the same field written in a circularly polarized basis.

For what concerns the first order in $\frac{1}{E_{p}}$, we will see now that it is not necessary to calculate $A^{(1)}$ and $B^{(1)}$, nor to give the explicit form of $\vec{ \mathcal E}^{(1)}$.
After propagating for a time $t$ the CMB field takes the form:
\begin{equation}
 \vec E(\vec r,t)=\Re\{\left[A \vec{ \mathcal E}_{+}e^{-i \omega_+ t}+B\vec{ \mathcal  E}_{-}e^{-i \omega_- t}\right]e^{i \vec p\cdot \vec r}\},
\end{equation}
where $\omega_\pm$ are defined in Eq. \eq{eq:disprel}.
We expand also $\omega_\pm$ in power of $\frac{1}{E_{p}}$, $\omega_\pm=\omega_0\pm\delta\omega$, with $\omega_0\equiv|\vec p|$ and $\delta\omega\equiv\frac{1 }{E_{p}} |\vec p|^2 \left( n_0 + \frac{{\vec n}\cdot {\vec p}}{|\vec p|}\right)^3$, so that the above expression for the field can be written up to the first order in $\frac{1}{E_p}$ as:
\begin{eqnarray}
 \vec E(\vec r,t)
&=&\Re\{\left[A_0 \vec{\mathcal E}_{+}^{(0)} (1-i \delta\omega \,t)+B_0\vec{ \mathcal  E}_{-}^{(0)}(1+i \delta\omega\, t)+\right. \\
&&+\left.\frac{1}{E_p} \left(A_1 \vec{ \mathcal E}_{+}^{(0)}+ B_1\vec{ \mathcal  E}_{-}^{(0)}+ A_0 \vec{ \mathcal E}_{+}^{(1)}+B_0 \vec {\mathcal E}_{-}^{(1)}\right)\right]e^{i (\vec p\cdot \vec r-\omega_0 t)}\}\nonumber \\
&=&\Re\{\left[\vec{ \mathcal E}- A_0 \vec{ \mathcal E}_{+} ^{(0)}i \delta\omega\, t+B_0\vec{ \mathcal  E}_{-} ^{(0)}i \delta\omega\, t\right]e^{i (\vec p\cdot \vec r-\omega_0 t})\}.\label{eq:propagationCMBphoton}
\end{eqnarray}
So up to the first order the field will be a combination of a term with the same polarization state $\vec{\mathcal{E}}$ as the field at time $t=0$, propagating with classical frequency $\omega_{0}$, and a term that describes the mixing of the zeroth-order components of the field. The polarization direction of the zeroth-order field is rotated during propagation, as it happens in presence of birefringence.

So we find that the first-order correction to the field, $\vec E^{(1)} $, is not significant within the orders of magnitude of our interest, and so we can consider the physical system as a linearly polarized field whose polarization direction rotates during propagation with angular velocity $\delta\omega$. In fact all the corrections to the zeroth-order field are of order $\frac{1}{E_p}$ as $\delta\omega$, but the magnitude of the terms with $\delta\omega$ can be amplified by the propagation time $t$, and this is indeed  the case for CMB photons. So all the corrections to the zeroth-order field are much less important than the corrections due to rotation of the zeroth-order linear polarization,  and we expect that, at least within the sensitivities presently reached by the instruments, we can at most see only the rotation of the zeroth-order linearly polarized transverse field:
\begin{eqnarray}
 \vec E(\vec r,t)&\simeq &\Re\left\{\left[\vec{ \mathcal E}^{(0)}(t=0)- A_0 \vec{ \mathcal E_{+}}^{(0)}i \delta\omega t+B_0\vec{\mathcal  E_{-}}^{(0)}i \delta\omega t\right]e^{i (\vec p\cdot \vec r-\omega_0 t})\right\}\nonumber\\
 &=&\Re\left\{\left[\left(\begin{array}{c}0\\a_1\\ a_2 \end{array}\right) -\frac{ a_2-i a_1}{\sqrt{2}} \left(
\begin{array}{c}
0  \\
  \frac{i}{ \sqrt{2}}   \\
 \frac{1}{\sqrt{2}}
\end{array}
\right)i \delta\omega t+\frac{a_2+i a_1}{\sqrt{2}}
        \left(
\begin{array}{c}
0  \\
 - \frac{i}{ \sqrt{2}}   \\
 \frac{1}{\sqrt{2}}
\end{array}
\right)i \delta\omega t\right]e^{i \tau_0}\right\}\nonumber\\
&=&\Re\left\{\left[\left(\begin{array}{c}0\\a_1\\ a_2 \end{array}\right) + \left(
\begin{array}{c}
0  \\
  a_2   \\
 -a_1
\end{array}
\right) \delta\omega t\right]e^{i \tau_0}\right\}\label{eq:CMBrotation}
\end{eqnarray}
where we have defined $\tau_{0}\equiv\vec p\cdot \vec r-\omega_0 t$.
The rotation angle $\alpha\equiv \delta\omega t$ is positive for rotations from the $\hat 2$ direction to the $\hat 1$ direction.

Another interesting feature emerging from this approximation is that in particular the longitudinal component of the field (which is univocally determined by imposing that Eq. \eq{eq:LinearCombination}) is constant with time, and is not amplified by the long propagation time.

\subsection{Anisotropic rotation of CMB Stokes parameters}

The relevant quantities that we can measure about CMB polarization are the Stokes parameters $Q$ and $U$ related to photons linear polarization properties.
Since they depend on the choice of reference frame,  in order to describe  CMB Stokes parameters all over the sky, we have to set a conventional all-sky reference frame.
This is of particular importance for us, since also the linear polarization rotation angle
 induced by (non-isotropic) birefringence can take opposite conventional signs depending on the reference frame choice.

 In the previous subsection we have conventionally set the rotation angle $\alpha(t,n_{0},\vec n\cdot\vec p,|\vec p|)$ to be positive if the rotation direction goes from the axis $\hat2\equiv\frac{-\vec p (\vec p\cdot \hat n)+\hat n |\vec p|^2}{|\vec p| \sqrt{p^2-(\hat n\cdot \vec p)^2}}$ to the axis $\hat 1\equiv \frac{ \hat n\times \vec p}{\sqrt{p^2-(\hat n\cdot \vec p)^2}}$ for a field coming toward us.

It is clear that, since we do not know the direction of $\vec n$, we cannot trivially change reference frame to the one conventional for CMB to the one, $\vec n$-dependent, that we have used up to now to describe non-isotropic photons propagation. But actually what really matters for the determination of the sign of $\alpha$ is the relative handedness of the two different reference frames, and the conventional reference frame of Eq. \eq{eq:baseBIS} that we have used to describe non-isotropic birefringence has an handedness which does not depend on the direction of $\vec n$.

 The reference frame in which conventionally the CMB Stokes parameters are defined at each point on the sky is a right handed coordinate system with $\hat z$ pointing outward toward the sky and ${\hat x, \hat y}$ directions defined as in appendix C.4 of \cite{Hinshaw:2003fc}.

In each point of the sky a rotation which in the reference frame \eq{eq:baseBIS} $\{\hat 1,\hat 2\}\equiv \{\frac{ \hat n\times \vec p}{\sqrt{p^2
 -(\hat n\cdot \vec p)^2}}, \frac{-\vec p (\vec p\cdot \hat n)
+\hat n |\vec p|^2}{|\vec p| \sqrt{p^2-(\hat n\cdot \vec p)^2}}\}$ used here to describe the transverse component of the electric field is from the $\hat 2$ direction to the $\hat 1$ direction (so that $\alpha$ is positive), will correspond in the CMB reference frame to a rotation from the $\hat y$ direction to the $\hat x$ one.


In the  reference frame on the sky  such that the observation direction is defined by the polar angles $\{\hat\theta,\hat\phi\}$, the direction of the vector $\vec n$ is given by the angles $\{\theta_n,\phi_n\}$. Then in the associated cartesian reference frame $\vec n$  can be expressed as
\begin{equation}
 \hat n=(\cos(\phi_n)\sin(\theta_n),\sin(\phi_n)\sin(\theta_n),\cos(\theta_n)).
\end{equation}
An analogous formula will hold for the photon propagation direction $\hat p$, which  is opposite to the observation direction  $\{\hat\theta,\hat\phi\}$, so
\begin{equation}
 \hat p=(-\cos(\phi)\sin(\theta),-\sin(\phi)\sin(\theta),-\cos(\theta))
\end{equation}
Then the dependence of the rotation angle $\alpha \equiv \delta\omega t=\frac{1 }{E_{p}} |\vec p|^2 \left( n_0 + \frac{{\vec n}\cdot {\vec p}}{|\vec p|}\right)^3 t$ on the observation direction takes the form\footnote{Note that here we are not taking into account the effect of photons energy redshift due to propagation in an expanding universe. To account for redshift  the expression for the rotation angle has to be changed in a way similar to the ones described in \cite{Gubitosi:2009eu}}:
\begin{equation}
\alpha(\theta,\phi) =\frac{1 }{E_{p}} |\vec p|^2 \left( n_0 - |\vec n|\left(\sin\theta\sin\theta_n\cos(\phi-\phi_n)+\cos\theta\cos\theta_n\right)\right)^3 t.\label{eq:alphaCMB}
\end{equation}

So at each point in the sky $\{\hat\theta,\hat\phi\}$, non-isotropic birefringence induced by the Lorentz violating Lagrangian of Eq. \eq{eq:lagrangianMP} that we are studying produces a mixing between the Stokes parameters, whose amount, for each given $\vec n$, depends on the observation direction and at the first order in the rotation angle $\alpha$ is given by:
\begin{eqnarray}
Q'(\theta,\phi)&=&Q(\theta,\phi)+2\alpha(\theta,\phi) U(\theta,\phi)\nonumber\\
U'(\theta,\phi)&=&U(\theta,\phi)-2\alpha(\theta,\phi) Q(\theta,\phi)\label{eq:StokesRotation}
\end{eqnarray}
Note that a positive rotation angle $\alpha$ here corresponds to a rotation from the $\hat y$ to the $\hat x$ direction.

\subsection{Consequences on  the polarization harmonic coefficients}
\label{sec:MultipoleCoeff}

In the case of isotropic birefringence the uniform rotation of Stokes parameters can be detected looking at the cross-correlation and auto correlation power spectra, since it produces a peculiar mixing between them \cite{Gubitosi:2009eu,Lepora:1998ix,Lue:1998mq,Feng:2006dp}.

In this subsection we will work out the modification to the full-sky power spectra induced by the direction-dependent rotation described above in this section, following the scheme presented in \cite{Gluscevic:2009mm,Kamionkowski:2008fp,Caldwell:2011pu}, but specializing the calculations to the case of our interest.
This will show that the kind of birefringence with anisotropic behavior described by Eq. \eq{eq:alphaCMB} cannot be detected using the standard tool of full-sky power spectra, so that it will be necessary to develop an alternative method of analysis for polarization data, which is done in the next section.

The rotation angle $\alpha$ is a scalar function defined on the sphere, so it can be expanded into spherical harmonics:
\begin{equation}
 \alpha(\theta,\phi)=\sum_{LM}\alpha_{LM}Y_{LM}(\theta,\phi) \label{eq:alphaExpansion}
\end{equation}
In the case of our interest, in which $\alpha$ takes the form \eq{eq:alphaCMB}, we need to sum only up to $L=3$, with the expansion coefficients given by (we have defined $\mathcal A\equiv \frac{|\vec p|^{2}}{E_{p}}t$):
\begin{eqnarray}
\alpha_{00}&= 2\,\mathcal A\, n_{0}\,(n_{0}^{2}+|\vec n|^{2})\sqrt{\pi}\\
\alpha_{10}&=- \mathcal A \, |\vec n|\, (5 n_{0}^{2}+|\vec n|^{2})\sqrt{3\pi}\cos{\theta_{n}}\\
\alpha_{1\pm1}&=\pm \frac{1}{5}\, \mathcal A  \,|\vec n| \,(5 n_{0}^{2}+|\vec n|^{2})e^{\mp i\phi_{n}}\sqrt{6\pi}\sin{\theta_{n}}\\
\alpha_{2 0}&= \mathcal A \, n_{0}\,|\vec n|^{2}\sqrt{\frac{\pi}{5}}\,(1+3\cos{2\theta_{n}})\\
\alpha_{2 \mp2}&= \mathcal A \, n_{0}\,|\vec n|^{2}e^{\pm 2i\phi_{n}}\sqrt{\frac{6\pi}{5}}\sin{\theta_{n}}^{2}\\
\alpha_{2 \mp1}&=\pm \mathcal A n_{0}|\vec n|^{2}e^{\pm i\phi_{n}}\sqrt{\frac{6\pi}{5}}\sin{\theta_{n}}\cos{\theta_{n}}\\
\alpha_{3 0}&=- \frac{1}{10}\, \mathcal A \,|\vec n|^{3}\sqrt{\frac{\pi}{7}}\,(3 \cos{\theta_{n}}+5\cos{3\theta_{n}})\\
\alpha_{3 \mp3}&=\mp\mathcal A\,  |\vec n|^{3}e^{\pm 3i\phi_{n}}\sqrt{\frac{\pi}{35}}\sin{\theta_{n}}^{3}\\
\alpha_{3 \mp 2}&=-\mathcal A\,  |\vec n|^{3}e^{\pm2i\phi_{n}}\,\sqrt{\frac{6 \pi}{35}}\sin{\theta_{n}}^{2}\cos{\theta_{n}}\\
\alpha_{3 \mp1}&=\mp \frac{1}{10}\, \mathcal A \, |\vec n|^{3}e^{\pm i\phi_{n}}\sqrt{\frac{3 \pi}{7}}\,(3+5\cos{2\theta_{n}})\sin{\theta_{n}}
\label{eq:alphaMultipoles}
\end{eqnarray}

We write the modification of multipoles of electric and magnetic modes of polarization  taking only the first order in $\alpha$ and considering the possibility of having
 primordial non-zero magnetic  modes.

Following the scheme presented in \cite{Gluscevic:2009mm,Kamionkowski:2008fp} we are able to evaluate the modification of the power spectra due to the polarization rotation:

\begin{eqnarray}
\label{eq:PowerSpectraMixing1}
 (C_\ell^{EB})'=&
 -\frac{2}{2\ell+1}\sum_m  \sum_{\stackrel{LM}{2\ell+L=even}}  \alpha_{LM}\left(C_\ell^{BB}-C_{\ell}^{EE} \right) H_{\ell\ell}^L\xi_{\ell m \ell m}^{LM}\\
 (C_{\ell}^{EE})'
 =& C_\ell^{EE}
\\
 ( C_{\ell}^{TB})'=&
 \frac{2}{2\ell+1}\sum_m\sum_{\stackrel{LM}{2\ell+L=even}}\alpha_{LM} C_{\ell}^{TE} H_{\ell\ell}^L\xi_{\ell m \ell m}^{LM}\\
( C_{\ell}^{TE})'=&C_\ell^{TE}
 \\
( C_{\ell}^{BB})'=& C_\ell^{BB}\label{eq:PowerSpectraMixing2}
\end{eqnarray}
where the coefficients $H_{\ell\ell}^L$ and the Wigner 3j-symbols $\xi_{\ell m \ell m}^{LM}$ are defined in \cite{Gluscevic:2009mm,Kamionkowski:2008fp}.
Notice that these expressions reduce to the ones  for the isotropic rotation case if $\alpha_{00}=\alpha_{0}$ and $\alpha_{LM}=0$ for all other $L,M$. In this case the mixing between the correlation power spectra are given by
\begin{eqnarray}\label{eq:PowerSpectraMixingIsotropic1}
 (C_\ell^{EB})'=&
 - 2 \alpha_{0}\left(C_\ell^{BB}-C_{\ell}^{EE} \right)\\
 (C_{\ell}^{EE})' =& C_\ell^{EE}
\\
 ( C_{\ell}^{TB})'=&2
\alpha_{0} C_{\ell}^{TE} \\
( C_{\ell}^{TE})'=&C_\ell^{TE}
 \\
( C_{\ell}^{BB})'=& C_\ell^{BB}\label{eq:PowerSpectraMixingIsotropic2}
\end{eqnarray}

Note that in this case a positive $\alpha_{0}$ corresponds to a rotation from the $\hat x$ axis to the $\hat y$ axis.

In the general case described by Eqs. \eq{eq:PowerSpectraMixing1}-\eq{eq:PowerSpectraMixing2}, all the correction terms for the spectra involve sums over $2\ell+L=even$, so take contributions only from $\alpha_{LM}$ with even $L$.  Moreover from the selection rules of the Wigner 3j-symbols, $\xi_{\ell m \ell m}^{LM}$ is different from zero only if $M=0$. So the only multipoles of $\alpha$ that contribute to the rotation of the power spectra are $\alpha_{L0}$, with $L$ even.

In our specific case this means that only the multipole coefficient $\alpha_{LM}$ with $\{L,M\}=\{0,0\},\{2,0\}$ contribute to modify the full-sky power spectra.
In particular, if the Lorentz-violating Lagrangian of Eq. \eq{eq:lagrangianMP} contains a purely space-like vector, $n_{\alpha}=(0,\vec n)$, which is the case that is here of primary interest, then the rotation angle \eq{eq:alphaCMB} reduces to
\begin{eqnarray}\label{eq:alphaCMBpureTIME}
\alpha(\theta,\phi)& =&\alpha_{max}\left(\sin\theta\sin\theta_n\cos(\phi-\phi_n)+\cos\theta\cos\theta_n\right)^3\\\nonumber
\alpha_{max}&=&-\frac{1 }{E_{p}} |\vec p|^2  |\vec n|^3 t
\end{eqnarray}
and from  Eq. \eq{eq:alphaMultipoles}, setting $n_{0}=0$, we see that  it needs only $L=1,3$ multipoles for the spherical harmonics expansion, so leaving no traces into the power spectra. An intuitive way to understand why this happens is to realize that, for $n_{\alpha}$ purely space-like, the rotation angle $\alpha$ takes opposite signs in opposite directions in the sky. Essentially, the modifications to the power spectra coming from one hemisphere of the sky compensate the modifications, of opposite sign, coming from the other hemisphere, and full-sky spectra are not affected by the polarization direction rotation. We concentrate on Lorentz violations induced by a purely-spacelike vector coupled with the electromagnetic field, since the standard analysis of full sky power spectra, that turned out to be useful in constraining the isotropic version of the model \cite{Gubitosi:2009eu}, cannot be exploited.
In the following section we introduce a new method of analysis, which indeed exploits the peculiarities of the model here of interest, and we apply it on  simulated data compatible with the ones expected from the \emph{Planck} satellite observation.

\section{Method of analysis}

The free parameters of the model that we want to constrain are the module and direction of the symmetry-breaking vector $\vec n$. Its module defines $\alpha_{max}$, the amount of rotation displayed by a photon coming from the direction of  maximum effect $\{\theta_{n}, \phi_{n}\}$ (in the antipodal direction the rotation is also maximum but has opposite sign). In order to avoid the  cancellation of the effect that would arise when working with (nearly) full sky CMB maps, we may work separately on the two hemispheres of the sky. However we don't know the direction $\vec n$ that would provide the natural division of the sky. A crucial observation to overcome this problem is that if we consider a small circular region of the sky centered at $\{\theta_{n},\phi_{n}\}$, then the power spectra estimated on the region are rotated according to Eqs. \eq{eq:PowerSpectraMixingIsotropic1}-\eq{eq:PowerSpectraMixingIsotropic2} by an angle\footnote{Here and in the following we indicate with $\alpha_{0}$ the effective rotation under the assumption of constant rotation, with $\alpha_{max}$ the maximum amount of rotation due to the anisotropic birefringence effect, and with the same symbols with a ( $\bar{}$ ) the corresponding estimated quantities }:


\begin{equation}
\alpha_{0}=A\times \alpha_{max}, \label{eq:aRESCALING}
\end{equation}
where  $A$ is given by the average of the function $ \left(\sin\theta\sin\theta_n\cos(\phi-\phi_n)+\cos\theta\cos\theta_n\right)^3$ over the region. For instance, if the angular radius of the region is $20^{\circ}$ we obtain $A=0.913$, while $10^{\circ}$ sets $A=0.978$. By taking this correction factor into account, the laws of transformation of the spectra
at first order in the rotation angle are ($\tilde C$ are the unrotated spectra):

\begin{eqnarray}
C_{\ell}^{TB}&=&2 \alpha_{0} \tilde C_{\ell}^{TE}\nonumber \\
C_{\ell}^{EB}&=&2\alpha_{0} (\tilde C_{\ell}^{EE}-\tilde C_{\ell}^{BB}).\label{eq:spectraDISK}
\end{eqnarray}
All the other spectra ($C_{\ell}^{TT}$, $C_{\ell}^{EE}$, $C_{\ell}^{BB}$ and $C_{\ell}^{TE}$) remain unchanged at linear order. Note that for $\alpha_0=0$ we expect $C_{\ell}^{TB}$ and $C_{\ell}^{EB}$ to be zero. We exploit this feature to constrain direction and amplitude of $\vec n$.
To this purpose we generate a set of  $1000$ masks to select, out of a CMB map, disks of angular radius $20^{\circ}$ \footnote{Note that we choose this value as a 
trade off between the goodness of the approximation in Eq. \eq{eq:aRESCALING} and the sample variance increase due to power spectrum estimation on very small regions of the sky.} and centers $\{\theta_{c},\phi_{c}\}$ distributed randomly in one half of the sky ($\theta_{c}\in[0^{\circ},90^{\circ}]$, $\phi_{c}\in[0^{\circ},360^{\circ}]$). In correspondence of each of these disks there is another one centered in the opposite direction. In this way we cover the entire available sky allowing for a certain degree of superposition among the disks.

We want to test the hypothesis that a disk is centered in the direction of maximum rotation. For this purpose we estimate the polarized power spectra out of each of them. We employ an estimator based on the pseudo spectra formalism  \cite{Polenta:2004qs,Hivon}, properly taking into account the effect of instrumental noise and incomplete sky coverage. According to Eq. \eq{eq:spectraDISK}, for each disk in one hemisphere we define the quantities:
\begin{eqnarray}\label{eq:spectraD_ell}
D^{TB}_{\ell} (\alpha_0)&=& C^{TB}_{\ell} - 2 \alpha_0 C^{TE}_{\ell}\\\nonumber
D^{EB}_{\ell} (\alpha_0) &=& C^{EB}_{\ell} - 2 \alpha_0 (C^{EE}_{\ell}-C^{BB}_{\ell})
\end{eqnarray}
using the measured power spectra ($C^{XY}_{\ell}$), under the assumption that the measured $EE$, $TE$ and $BB$ spectra are unchanged at first order in the rotation angle (see Eqs. \eq{eq:spectraDISK}).
Since in the opposite hemisphere the effect is expected to be identical except for a sign flip in $\alpha_0$, the estimators $\bar{D}^{TB}_{\ell}$ and $\bar{D}^{EB}_{\ell}$ for the antipodal disks can be easily built just changing the sign of $\alpha_0$ in Eq. \eq{eq:spectraD_ell}. In order to determine the direction of maximum rotation we compute, for each region $i$ ($i=1,...,1000$) and its antipodal one, the $\alpha_0=\bar\alpha_{0i}$ which minimizes the $\chi^{2}(\alpha_0)$ given by:
\begin{eqnarray}\label{eq:chisq}
\chi^{2}(\alpha_0)&=&\sum_{\ell\ell'}D^{TB}_{\ell} \Sigma_{TB,\ell\ell'}^{-1}D^{TB}_{\ell'} + \sum_{\ell\ell'}D^{EB}_{\ell} \Sigma_{EB,\ell\ell'}^{-1}D^{EB}_{\ell'}+\\\nonumber
&+&\sum_{\ell\ell'}\bar{D}^{TB}_{\ell} \bar{\Sigma}_{TB,\ell\ell'}^{-1}\bar{D}^{TB}_{\ell'}+\sum_{\ell\ell'}\bar{D}^{EB}_{\ell} \bar{\Sigma}_{EB,\ell\ell'}^{-1}\bar{D}^{EB}_{\ell'},
\end{eqnarray}
where $\Sigma_{XY,\ell\ell'}$ ($\bar{\Sigma}_{XY,\ell\ell'}$) are the covariance matrices of the $D^{XY}_{\ell}$ ($\bar{D}^{XY}_{\ell}$) estimators. 
We end up with an estimate of $\alpha_0$, i.e. $\bar\alpha_{0i}$, for each region and we project these values onto a sky map, taking the average of the $\bar\alpha_{0i}$ values associated to overlapping regions. Our estimate of $\alpha_{max}$, i.e. $\bar\alpha_{max}$, is the maximum value of the $\bar\alpha_0$ map rescaled by the correction factor A as given in Eq. \eq{eq:aRESCALING}. The uncertainties on $\bar\alpha_{max}$ can be assessed via Monte Carlo (MC) simulations, as it will be explained in detail in the next section.
The final step of the analysis is to derive an estimate of the direction of the symmetry breaking vector, $\{\bar\theta_{n},\bar\phi_{n}\}$, and the confidence intervals associated to it. By testing the method on simulations, it turns out that we cannot evaluate simultaneously $\bar\alpha_{max}$ and $\{\bar\theta_{n},\bar\phi_{n}\}$ by minimizing a joint $\chi^{2}(\alpha_0,\theta,\phi)$, due to the degeneracies between the three parameters. For this reason we take the direction of the maximum value of the $\bar\alpha_0$ map as the best-fit of $\{\theta_{n},\phi_{n}\}$ and we derive the uncertainty on this estimate by slicing the function $\bar\alpha_{0}(\theta,\phi)$ in correspondence of the $1$ and $2\sigma$ errors on $\bar\alpha_{0}$.

\section{Forecasts for \emph{Planck}}

\begin{figure}[!htb]
\begin{center}
\includegraphics[scale=0.35,angle=90]{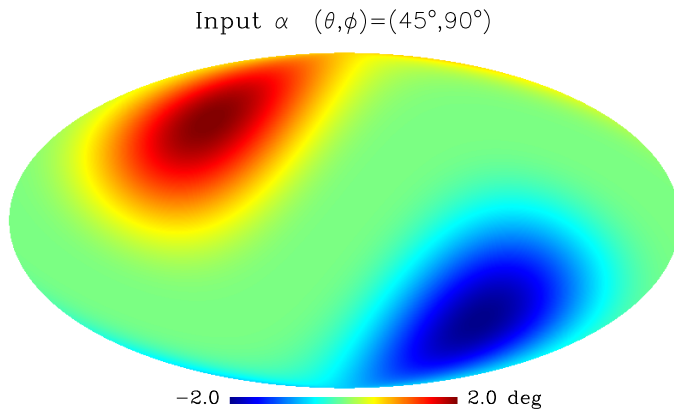}\label{Q_rot_45_90}
\caption{Map of the birefringence angle (see Eq. \eq{eq:alphaCMBpureTIME}) as used in the simulations. The map has been generated in the Galactic coordinates, using the Healpix scheme \cite{Gorski} with resolution parameter $N_\mathrm{side}=1024$.}
\label{fig:Diff_maps}
\end{center}
\end{figure}

In this section we derive forecasts to constrain the anisotropic birefringence model set forth in Section~\ref{sec:birefringence} with \emph{Planck} \cite{ref:Planck_mission}, by applying the method outlined above on simulated data.
In particular, we draw (T,Q,U) CMB simulated maps as Gaussian realizations of the WMAP best-fit cosmological model \cite{Komatsu:2010fb}, only accounting for pure scalar perturbations ($\tilde C_{\ell}^{BB}=0$). We then introduce an anisotropic rotation of the Stokes parameters according to Eq. \eq{eq:StokesRotation}, with the rotation angle given by Eq. \eq{eq:alphaCMBpureTIME} where we fix $\alpha_{max}=1.98^\circ$ while $\{\theta_{n},\phi_{n}\}=\{45^{\circ} ,90^{\circ}\}$. Angles are given in the Galactic coordinate system with $\theta \in [-90^{\circ},+90^{\circ}]$ and $\phi \in [0^{\circ}, 360^{\circ}]$. Furthermore, note that the value chosen for $\alpha_{max}$ is consistent with presently available upper limits on the isotropic version of the birefringence model~\cite{Gubitosi:2009eu,Komatsu:2010fb,Wu:2008qb, Xia:2009ah}.
In figure \ref{fig:Diff_maps} we report the map of the birefringence angle (see Eq. \eq{eq:alphaCMBpureTIME}) as input to the simulations. The map points out the symmetries expected for the proposed birefringence model, and in fact we have verified that, averaging over $1000$ simulations, the full-sky power spectra of the rotated maps are consistent with those of the unrotated ones.



\begin{figure}[!htb]
\begin{center}
\includegraphics[scale=0.35,angle=90]{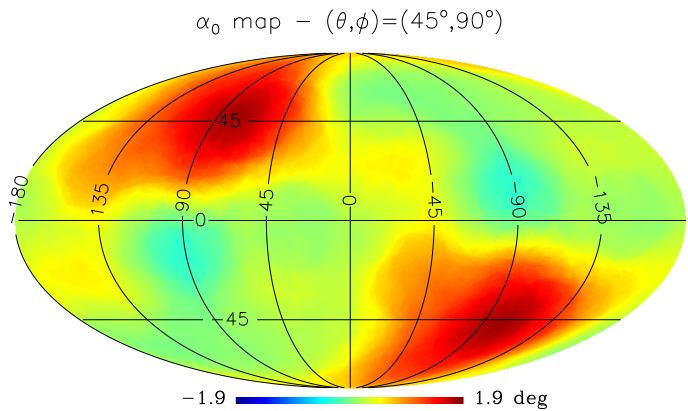}
\caption{Distribution on the sphere of the estimated $\bar\alpha_{0}$. It has been obtained by analyzing CMB plus noise simulations with a birefringence angle~\eq{eq:alphaCMBpureTIME} set by $\alpha_{max}$ is $1.98^{\circ}$ and $\{\theta_{n},\phi_{n}\}=\{45^{\circ},90^{\circ}\}$. Note that here we are plotting the absolute value of the rotation angle $\bar\alpha_{0}$. The map has been generated in the Galactic coordinates of the Healpix scheme \cite{Gorski} with the resolution parameter $N_{side}=1024$.
}
\label{fig:maps_alpha}
\end{center}
\end{figure}

We add to the rotated CMB maps described above  isotropic white noise based on the \emph{Planck} $143$ GHz channel sensitivity~\cite{Lamarre:2010}, assuming a 30 months long mission. We choose this particular frequency channel because it is the one with the highest signal to noise ratio in polarization at high resolution. By following the procedure described in the previous section, we apply our set of masks to the simulated maps, computing for each mask the power spectra and deriving the best-fit of $\alpha_0$ by minimizing the $\chi^2$ in Eq. \eq{eq:chisq}. In figure~\ref{fig:maps_alpha} the recovered values of $\alpha_0$ are projected onto a map and, as one may see, they trace a smooth distribution with the highest values concentrated, as expected, around the input values of $\{\theta_{n},\phi_{n} \}$.

\begin{figure}[!htb]
\begin{center}
\includegraphics[scale=0.45]{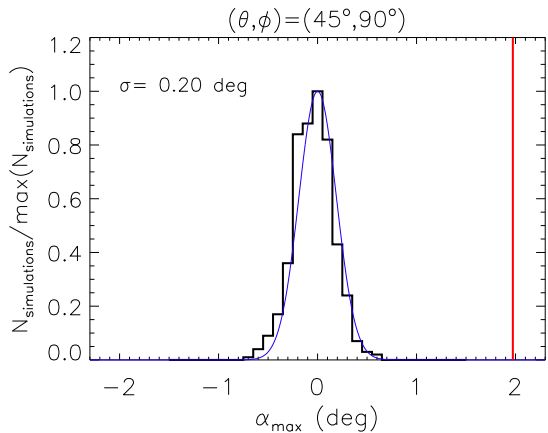}
\caption{Histogram of $\bar{\alpha}_{max}$ estimated for 500 MC simulations of unrotated CMB plus noise on the mask where the rotation effect is maximum. The solid curve in the plots represents the $\chi^2(\alpha)$ derived from the standard analysis on the same maximum effects mask, while the $\sigma$ in the label is the standard deviation of the values in the histogram. The plotted quantities have been rescaled by the factor $\sqrt{(\mathbf{W}^T\mathbf{C}^{-1}\mathbf{W})^{-1}}$ given in Eq. \eq{eq:scale_sigma}. The vertical line is the $\bar{\alpha}_{max}$ evaluated from maps where the birefringence effect is present.}
\label{fig:histogram_alpha}
\end{center}
\end{figure}

\begin{figure}[!htb]
\begin{center}
\includegraphics[width=0.35\textwidth]{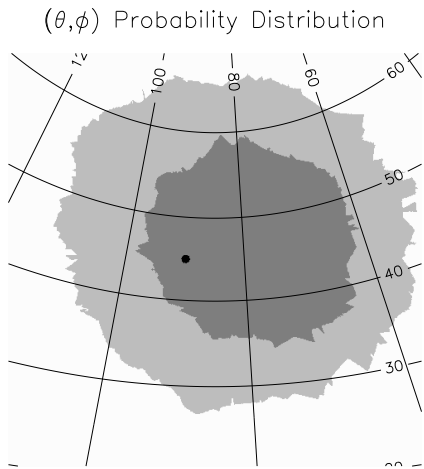}
\caption{Contour plot of the confidence intervals on $\{\bar\theta_{n},\bar\phi_{n}\}$, obtained by slicing the function $\bar\alpha_{0}(\theta,\phi)$ in correspondence of the $1$ and $2\sigma$ errors on $\bar\alpha_{0}$. The black dot marks the input $\{\theta_{n},\phi_{n}\}$.}
\end{center}
\label{fig:surface}
\end{figure}

As described in the previous section, we can estimate the best-fit $\bar{\alpha}_{max}$ from the maximum of this map. We derive the uncertainties on $\bar{\alpha}_{max}$ by means of 500 MC simulations of unrotated CMB plus noise, the same simulations we use to assess the covariance matrices of the power spectra ($C^{XY}_{\ell}$). Note that considering unrotated CMB maps is just an approximation, nevertheless we expect that the contribution of the rotation itself to the power spectra covariances is subdominant. For the analyzed maps with the anisotropic rotation, we can individuate the disk $i$ where the birefringence effect is maximum and compute the best-fit $\bar\alpha_{0i}$ for each of the MC simulations on that specific mask.
Using these 500 values of $\bar{\alpha}_{0i}$ we can trace a ``frequentist" probability distribution for the parameter and evaluate the one sigma error as the standard deviation of these values ($\sigma_S$). However, when we build the $\bar\alpha_{0}$ map as in figure~\ref{fig:maps_alpha}, several, partially overlapping, disks contribute to constrain $\bar\alpha_0$, other than the particular disk $i$ where the estimates is found to be maximal. Estimates from these disks contribute to lower the variance associated to $\bar\alpha_0$ with respect to $\sigma_S$, but they are obviously correlated due to overlap. In order to estimate the latter variance, we model the correlation degree in the overlapping disks as the fractional overlapping area between each pair of contributing disks. In other words, this correlation is 0 if two disks are totally disjointed, and 1 in the limit of perfectly overlapping disks, all intermediate values being allowed. We then take as one sigma error:

\begin{equation}
\label{eq:scale_sigma}
\sigma=\sigma_S \sqrt{(\mathbf{W}^T\mathbf{C}^{-1}\mathbf{W})^{-1}},
\end{equation}

where $\mathbf{W}$=[1,...,1] is a design matrix, while $\mathbf{C}^{-1}$ is the correlation matrix between $\bar\alpha_{0i}$ of different overlapping masks computed as described above. (Note how the rescaling factor reduces to the well familiar $1/\sqrt{N}$ in the unrealistic case of $N$ disjointed disks.) In figure~\ref{fig:histogram_alpha} we present the histogram that traces the probability distribution of $\bar\alpha_{max}$, as derived from the 500 MC simulations, and superimposed to it there is also the $\chi^2(\alpha)$ curve obtained from the standard analysis of Eq. \eq{eq:chisq} on the considered maximum effect mask (both rescaled by the factor $\sqrt{(\mathbf{W}^T\mathbf{C}^{-1}\mathbf{W})^{-1}}$ given in Eq. \eq{eq:scale_sigma}). The width of the histogram is compatible with that of the $\chi^2(\alpha)$ curve, confirming that neglecting the rotation in the MC simulations has a negligible impact. We want to stress that from figure~\ref{fig:histogram_alpha} is evident that we can detect the birefringence effect (vertical red line) at high significance. Once the uncertainty on $\bar{\alpha}_{max}$ has been assessed, we can derive the confidence intervals on $\{\bar{\theta}_{n},\bar{\phi}_{n}\}$ (see figure~\ref{fig:surface}) as explained at the end of the previous section, i.e. by slicing the function $\bar\alpha_{0}(\theta,\phi)$ in correspondence of the $1$ and $2\sigma$ errors on $\bar\alpha_{0}$ (which are related to the uncertainties on $\alpha_{max}$ through Eq. \eq{eq:aRESCALING}).


Our main results are summarized in Table~\ref{tab:results_val}, showing that the analysis recovers quite well the input parameters of the simulations. This means that our method of analysis is able to detect an anisotropic rotation effect (with $\alpha_{max}$ given by the current upper limits on the isotropic case) and give reliable constraints on its parameters for data which are compatible with the \emph{Planck} sensitivity.

\begin{table}[!htb]
\begin{center}
\begin{tabular}{cccc}
$\bar\alpha_{max}\pm \Delta\bar\alpha_{max}$ & $\Delta\bar\theta_{n}$ & $\Delta\bar\phi_{n}$\\
\hline
\hline
\hline
$1.97 \pm 0.20$ &$[23.9,63.0] $ &$[49.5,114.7]$ \\
\end{tabular}
\caption{Estimates obtained for the three parameters of the model (all expressed in degrees). $\Delta\bar\theta_{n}$ and $\Delta\bar\phi_{n}$ have been derived at 2 $\sigma$ of $\bar\alpha_{max}$. We recall that the input value for $\alpha_{max}$ is $1.98^{\circ}$ and $\{\theta_{n},\phi_{n} \} = \{45^\circ, 90^\circ \}$.}
\label{tab:results_val}
\end{center}
\end{table}


In order to check for the robustness of the analysis, we perform further consistency tests. For instance, we verify that results remain stable by (reasonably) changing the range of multipoles taken into account in Eq. \eq{eq:chisq}. In particular, results reported in Table~\ref{tab:results_val} and in the previous figures have been obtained by considering $\ell \in [100,2000]$, where $\ell_{min}$ is bound by the fact that larger scales are not properly constrained by power spectra computed on small regions of the sky, while $\ell_{max}$ corresponds roughly to the smallest angular scale that can be recovered at the 20\% of the instrumental beam transfer function (which, for the \emph{Planck} $143$ GHz channel, we assume to be a Gaussian with FWHM=$7.1'$~\cite{Lamarre:2010}). Moreover we derive constraints of the three parameters of the model by analyzing separately TB and EB rotated spectra. As foreseable, constraints obtained from EB spectra are tighter, since the cosmic variance of these spectra, which we recall have been computed on small patches of the sky, is smaller than that of TB spectra. Still for checking purposes, we have analysed the two hemispheres of the rotated maps separately, obtaining compatible estimates of the parameters. Results of all these checks are reported in Table~\ref{tab:results_checks}.

\begin{table}[!htb]
\begin{center}
\begin{tabular}{cccc}
& $\bar\alpha_{max}\pm \Delta\bar\alpha_{max}$& $\Delta\bar\theta_{n}$ & $\Delta\bar\phi_{n}$\\
\hline
\hline
\hline
TB & $2.38\pm 0.34$ &$ [21.9,72.2] $ &$ [33.1, 137.8] $\\
EB & $1.83\pm 0.22$ &$[25.2, 70.7]$ &$[49.8, 113.6] $\\
N & $2.25\pm 0.28$ &$[22.3, 66.3] $ &$[59.3, 112.3]$\\
S & $1.89 \pm 0.29$ &$[25.2, 73.6] $ &$[27.5, 135.2] $\\
\end{tabular}
\caption{Consistency checks on the estimated model parameters. The first two rows refer to the results obtained analysing $D^{TB}_{\ell}$ and $D^{EB}_{\ell}$ spectra separately. The last two rows, instead, show results derived on a single hemisphere (respectively N=North and S=South). $\Delta\bar\theta_{n}$ and $\Delta\bar\phi_{n}$ have been derived at 2 $\sigma$ of $\bar\alpha_{max}$. All values reported in this table are expressed in degrees. We recall that the input value for $\alpha_{max}$ is $1.98^{\circ}$ and $\{\theta_{n},\phi_{n} \} = \{45^\circ, 90^\circ \}$. The results of the full analysis are reported in Table 1.}
\label{tab:results_checks}
\end{center}
\end{table}

%
A natural continuation of this analysis is to apply it to presently
available data, such as those collected by WMAP, and we will address
this point in a future paper.


\section{Conclusions}

It was recently shown \cite{Gubitosi:2009eu} that data from CMB polarization observations can be used to constrain quantum-gravity effects producing anomalous light propagation, with a sensitivity that is sufficient to test effects originating at the Planck scale. However, these observations were thought to be able to provide limits that are not really competitive with the ones coming from astrophysics,  due mainly  to the lower energy characterizing CMB photons.

We established here
that CMB data can also provide powerful constraints on non-isotropic effects, since they provide information from radiation coming from almost all directions on the sky. And in this case they can be truly competitive with the corresponding constraints on non-isotropic anomalous light propagation obtainable from observations of
astrophysical sources, since in astrophysics one is limited to gaining information
 on only a few directions of propagation of signals (see also the discussion in \cite{anysotropy}).
 
For the specific  non-isotropic anomalies for light propagation we considered here,
 which are expected in the  quantum gravity-inspired models of Ref.~\cite{Myers:2003fd,anysotropy},
  we found that they  don't produce any signature in the  CMB full-sky power spectra,
  due to cancellations induced by the
peculiar symmetries of the effect. Nevertheless, the anomalous effects can be
anyway detected through a method that still exploits the power spectra
formalism while operating on small patches of the sky in order avoid
the cancellations.
  

The effect we studied produces a rotation of the CMB polarization direction (birefringence), whose amount depends on the observation direction.
We characterized the effect in terms of the amount of rotation in the direction of maximum effect ($\alpha_{max}$) and the two angles identifying this direction ($\theta_{n}$ and $\phi_{n}$). These parameters are related in a simple way to the parameters characterizing the Lagrangian describing our model (see Eq. \eq{eq:alphaCMBpureTIME}).

We have shown that data coming from the \emph{Planck} satellite will be able to constrain anomalous non-isotropic light propagation with a sensitivity of $0.2$ degrees on the amount of  rotation in the direction of maximum effect and will also allow to identify the special direction pointed by the symmetry breaking vector with an uncertainty of roughly $40$ degrees on the $\theta$ angle and $60$ degrees on the $\phi$ angle.

The sensitivity on $\alpha_{max}$ translates into a sensitivity on the parameter that indicates how far the constraint is from the Planck scale, \emph{i.e.} the module of the symmetry-breaking vector $\vec n$. We expect this parameter to be of order one if the effect is generated at the Planck scale.

Exploiting the relation in Eq. \eq{eq:alphaCMBpureTIME}, and taking into account also the correction due to photon redshift as described in \cite{Gubitosi:2009eu}, the sensitivity on $\alpha_{max}$ can be translated into a sensitivity on $|\vec n|$ of the order of  $10^{-1}$, which is one order of magnitude better than the level required to test the Planck scale.



Note also that the sensitivity can be further improved exploiting the availability of different energy channels in the \emph{Planck} observations and the predicted dependence of the effect on the square of the photons energy.

\ack

Part of the research of LP was carried out at the Jet Propulsion Laboratory, California Institute of Technology, under a contract with the National Aeronautics and Space Administration.\\
This work is supported by PRIN-INAF, "Astronomy probes fundamental physics". Support was given by the Italian Space Agency through the ASI contracts Euclid-IC (I/031/10/0).\\
This research used resources at NERSC, supported
by the DOE under Contract No. DE-AC03-76SF00098, and
at CASPUR (Rome, Italy: special thanks are due to M.
Botti and F. Massaioli). \\We also acknowledge support
from Contract Planck LFI activity of Phase E2.

\section*{References}
\bibliographystyle{iopart-num}
\bibliography{bibliography}

\end{document}